\documentclass{article}

\usepackage[nonatbib, final]{neurips_2021}
\usepackage[utf8]{inputenc} % allow utf-8 input
\usepackage[T1]{fontenc}    % use 8-bit T1 fonts
\usepackage{hyperref}       % hyperlinks
\usepackage{url}            % simple URL typesetting
\usepackage{bm}
\usepackage{booktabs}       % professional-quality tables
\usepackage{amsfonts}       % blackboard math symbols
\usepackage{amssymb}
\usepackage{nicefrac}       % compact symbols for 1/2, etc.
\usepackage{microtype}      % microtypography
\usepackage{xcolor}         % colors
\usepackage{graphicx} % add figures
\usepackage{float} % allow image anchor "H"
\usepackage{caption}
\usepackage{subcaption}

\usepackage[style=ieee]{biblatex} % ieee bib style
\usepackage{xcolor}

\title{Investigation of Independent Reinforcement Learning Algorithms in Multi-Agent Environments}

\author{%
  Ken Ming Lee \\
  University of Waterloo\\
  Waterloo, Canada \\
  \texttt{km23lee@uwaterloo.ca} \\
  \And 
  Sriram Ganapathi Subramanian\\
  University of Waterloo\\
  Waterloo, Canada \\
  \texttt{s2ganapa@uwaterloo.ca} \\
  \And
  Mark Crowley\\
  University of Waterloo\\
  Waterloo, Canada \\
  \texttt{mcrowley@uwaterloo.ca} \\

}

\begin{document}

\maketitle

\begin{abstract}
Independent reinforcement learning algorithms have no theoretical guarantees for finding the best policy in multi-agent settings. However, in practice, prior works have reported good performance with independent algorithms in some domains and bad performance in others. Moreover, a comprehensive study of the strengths and weaknesses of independent algorithms is lacking in the literature. In this paper, we carry out an empirical comparison of the performance of independent algorithms on four PettingZoo environments that span the three main categories of multi-agent environments, i.e., cooperative, competitive, and mixed. We show that in fully-observable environments, independent algorithms can perform on par with multi-agent algorithms in cooperative and competitive settings. For the mixed environments, we show that agents trained via independent algorithms learn to perform well individually, but fail to learn to cooperate with allies and compete with enemies. We also show that adding recurrence improves the learning of independent algorithms in cooperative partially observable environments.
\end{abstract}

\section{Introduction}
One of the simplest ways to apply reinforcement learning in multi-agent settings is to assume that all agents are independent of each other. In other words, every other agent is seen as part of the environment from any agent's perspective. Independent algorithms (i.e., single-agent algorithms) face the issue of non-stationarity in the multi-agent domain due to the violation of the Markovian property in a Markov Decision Process \cite{NIPS1999_e8d92f99}. As a result, independent algorithms lack convergence guarantees, and are not theoretically sound in the multi-agent setting 
\cite{tan1993multi}. 
Despite these shortcomings, independent algorithms have the advantage of requiring lower computational resources and being easier to scale to large environments than traditional multi-agent algorithms which perform exact opponent modelling of each agent. In practice, prior works have reported mixed performance for independent algorithms in different multi-agent domains \cite{zawadzki2014empirically, 10.1371/journal.pone.0172395, shoham2008multiagent, openai2019dota, coma, lowe_maddpg, qmix}. However, a study of the strengths and weaknesses of independent algorithms across various categories within the multi-agent domain is lacking in the literature.

In this paper, we investigate the empirical performance of independent algorithms in multi-agent settings, and compare them to various multi-agent algorithms under the Centralized Training and Decentralized Execution scheme \cite{KRAEMER201682, ctde}. We evaluate these algorithms on 4 multi-agent environments from the PettingZoo library \cite{terry2020pettingzoo}, which span the 3 main categories of multi-agent environments (i.e., cooperative, competitive and mixed) \cite{original_marl_taxonomy, app11114948, zhang2021multiagent, gronauer2021multi}. We show that independent algorithms can perform on par with multi-agent algorithms in the cooperative, fully-observable setting, and adding recurrence allows them to perform well compared to multi-agent algorithms in partially observable environments. In the competitive setting, we show that parameter sharing alongside the addition of agent indicators allow independent algorithms to outperform some multi-agent algorithms, such as Multi-Agent Proximal Policy Optimization \cite{yu2021mappo}, and Multi-Agent Deep Deterministic Policy Gradient \cite{lowe_maddpg}, in fully-observable environments. For the mixed setting, we show that agents of independent algorithms learn to perform well individually, but fail in learning to cooperate with allies and compete against enemies.

\section{Background Information}
In this section, we provide readers with a brief overview of the various concepts and algorithms that are used throughout the paper.
\subsection{Reinforcement Learning}
In Reinforcement Learning (RL), an agent interacts with the environment by making sequential decisions \cite{sutton2018reinforcement}. At every time step, denoted as $t$, the agent observes a state $s_t$ from the environment, and takes an action $a_t$. This action is executed in the environment, which returns a reward $r_t$ and the next state $s_{t+1}$ that are determined by the reward function $R(s_t, a_t)$ and the transition probability, $P(s_{t+1}|s_t, a_t)$, respectively. Critically, $R(s_t, a_t)$ and $P(s_{t+1}|s_t, a_t)$ are part of the environment, and are usually unknown to the agent of a model-free RL algorithm. Since the transition probability $P(s_{t+1}|s_t, a_t)$ conditions the next state $s_{t+1}$ purely on the current state $s_t$ and action $a_t$, it satisfies the Markov property \cite{markov1954theory}. This interaction between the agent and the environment is called a Markov Decision Process (MDP) \cite{mdp}. The objective of an RL agent is to learn a policy $\pi(a_t|s_t)$, which maps a state to an action that maximizes the expected cumulative reward it receives from the environment. This is written as $\sum_t\, \gamma^t r_t$, where $\gamma \in [0,1)$ represents a discount factor on future rewards.

\subsection{Multi-Agent Reinforcement Learning}
The single-agent MDP framework is extended to the Multi-Agent Reinforcement Learning (MARL) setting in the form of stochastic games \cite{stochastic_games}. In an $N$-agent stochastic game, at every time step, each of the $n$ agents, identified by $j \in \{1, 2, \ldots, n\}$ across all agents, takes an action $u_t^j$. The joint action $ \boldsymbol{u_t} \triangleq \{u^1_t, \ldots, u^N_t \}$ determines the rewards obtained by each agent. State transitions of the environment are determined by the transition probability $P(s_{t+1} |s_t, u_t)$, which conditions on the state and the joint action at timestep $t$.

\subsection{Centralized Training and Decentralized Execution}
While it is technically possible to learn a centralized controller that maps a state to a distribution over the joint action space, the number of possible combinations of actions grows exponentially with the number of agents. This makes centralized control intractable for environments with many agents. Therefore, this paper is mainly focused on multi-agent algorithms which correspond to a Centralized Training and Decentralized Execution (CTDE) scheme \cite{KRAEMER201682, ctde}. A CTDE algorithm has two phases. During the control phase, where policies are deployed in the environment, rather than using a centralized controller to take actions for all agents, decentralized agents make decisions based on their individual observations. During the prediction phase, centralized training is performed, which allows for extra information (e.g. the state) to be utilized, as long as this is not required during the control phase.

\subsection{Cooperative, Competitive and Mixed}

This paper follows the convention of classifying every multi-agent algorithm and environment studied into one of three categories -- cooperative, competitive, or mixed (cooperative-competitive) \cite{original_marl_taxonomy, app11114948, zhang2021multiagent, gronauer2021multi}.  

In the cooperative setting, agents collaborate with each other to achieve a common goal. As a result, it is very common for all agents to share the same reward function (i.e., a team goal) \cite{NIPS2003_c8067ad1}. Also known as the multi-agent credit assignment problem, every agent has to deduce its own contributions from the team reward \cite{NIPS2003_c8067ad1}. Algorithms studied in this paper that explicitly address the multi-agent credit-assignment problem include QMIX \cite{qmix} and Counterfactual Multi-Agent Policy Gradients (COMA) \cite{coma}. Additionally, the CommNet \cite{commnet} extension on top of COMA is utilized for specific cooperative environments. Other multi-agent algorithms that are considered for the cooperative scenario include Multi-Agent Deep Deterministic Policy Gradient (MADDPG) \cite{lowe_maddpg} and Multi-Agent Proximal Policy Optimization (MAPPO) \cite{yu2021mappo}.

In the competitive setting, agents play a zero-sum game, where one agent's gain is another agent's loss. In other words, $\sum_a r(s, u, a) = 0\, \forall s, u$. Algorithms that are studied specifically in this paper include Deep Reinforcement Opponent Network (DRON) \cite{dron}, MADDPG and MAPPO. MADDPG and MAPPO  learn a separate critic for every agent, which gives the algorithms flexibility to learn different behaviours for agents with different reward functions.

In a mixed or cooperative-competitive setting, environments are neither zero-sum (competitive) nor cooperative, and they do not necessarily need to be general-sum either. A common setting would be environments that require every agent to cooperate with some agents, and compete with others \cite{original_marl_taxonomy, app11114948, zhang2021multiagent}. MADDPG and MAPPO are used here for the same reason as the competitive setting.

\subsection{Independent Algorithms and Non-Stationarity}
One naive approach for applying single-agent RL to the multi-agent setting would be the use of independent learners, where each agent treats every other agent as part of the environment, and learns purely based on individual observations. In a multi-agent setting, the transition probability $P$ and reward function $R$ are conditioned on the joint action $u$. Since all agents in the environment are learning, their policies constantly change. Therefore, from each independent learner's perspective, the transition probability and reward function appear non-stationary, due to the lack of awareness of other agents' actions. This violates the Markovian property of an MDP, which then causes independent algorithms to be slow to adapt to other agents' changing policies, and as a result, face difficulties in converging to a good policy \cite{papoudakis_nonstationarity, non_stationarity_survey, dron}. 

In this paper, we chose to use a popular off-policy algorithm, Deep Q-Network (DQN) \cite{mnih2015dqn}, and an on-policy algorithm, Proximal Policy Optimization (PPO) \cite{ppo}. In specific partially observable environments, Deep Recurrent Q-Network (DRQN) \cite{drqn} is also utilized. Following the work of Gupta et al. \cite{param_sharing}, parameter sharing is utilized for all independent algorithms, such that experiences from all agents are trained simultaneously using a single network. This allows the training to be more efficient \cite{param_sharing}. The aforementioned independent algorithms are tested in all 3 categories of multi-agent environments. 

\section{Experimental Setup}
In this section, we introduce the environments used for the experiments, specify the various preprocessing that were applied, and other relevant implementation details. 
\subsection{Environments}
\begin{figure}
     \centering
     \begin{subfigure}[t]{0.15\textwidth}
        \vskip 0pt
         \includegraphics[width=\textwidth]{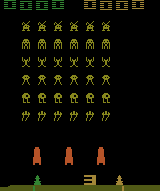}
         \caption{Space Invaders (Atari, cooperative)}
         \label{fig:space_invaders}
     \end{subfigure}
     \qquad
     \begin{subfigure}[t]{0.2\textwidth}
        \vskip 0pt
         \frame{\includegraphics[width=\textwidth]{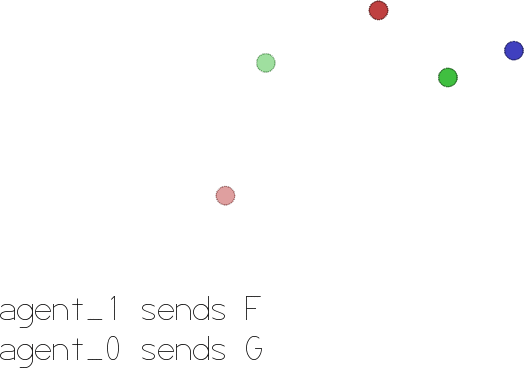}}
         \vspace{4pt}
         \caption{Simple Reference (MPE, cooperative)}
         \label{fig:simple_reference}
     \end{subfigure}
    \qquad
    \begin{subfigure}[t]{0.15\textwidth}
        \vskip 0pt
         \includegraphics[width=\textwidth]{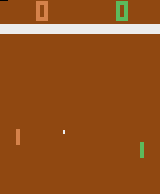}
         \caption{Pong (Atari, competitive)}
         \label{fig:pong}
     \end{subfigure}
     \qquad
     \begin{subfigure}[t]{0.17\textwidth}
        \vskip 0pt
        \frame{\includegraphics[width=\textwidth]{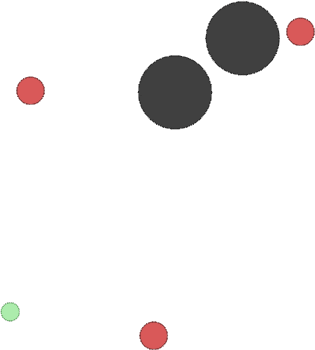}}
         \caption{Simple Tag (MPE, mixed)}
         \label{fig:simple_tag}
     \end{subfigure}
        \caption{The four PettingZoo environments used in the experiments. All figures were obtained from \url{https://pettingzoo.ml/}}
        
\end{figure}
The experiments were performed on multiple multi-agent environments from the PettingZoo library \cite{terry2020pettingzoo}, which contains the Multi-Agent Particle Environments (MPE) \cite{lowe_maddpg, mpe} and multi-agent variants of the Atari 2600 Arcade Learning Environment (ALE) \cite{bellemare13arcade, terry2020arcade}. 

For the cooperative setting, experiments were performed on a modified version of the 2-player Space Invaders \cite{bellemare13arcade, terry2020arcade}, and the Simple Reference MPE environment \cite{lowe_maddpg, mpe}. In Space Invaders (Fig.~\ref{fig:space_invaders}), both agents share the common goal of shooting down all aliens. To make Space Invaders cooperative, we removed the (positive) reward that is given to a player whenever the other player gets hit. Additionally, the environment rewards every agent individually by default. Since a number of cooperative multi-agent algorithms (e.g., QMIX and COMA) assume that only a team reward is given, we modified the reward function such that a team reward is given instead (i.e., both agents receive the sum of their individual rewards). This setup exemplifies the multi-agent credit assignment problem, the effect of which is studied more closely in the Section~\ref{section:effect_of_team_reward}. 
On the other hand, in the Simple Reference environment (Fig.~\ref{fig:simple_reference}), two agents are rewarded by how close they are to their target landmark. However, the target landmark of an agent is only known by the other agent, as a result communication is required for both agents to navigate successfully to their target landmarks.   

For the competitive setting, we performed experiments on the 2-player variant of the original Atari Pong environment (Fig.~\ref{fig:pong}). For the mixed setting, we opted for the Simple Tag MPE environment (Fig.~\ref{fig:simple_tag}), which is a Predator-Prey environment \cite{mpe}. This environment consists of 4 agents -- 3 predators and a prey. The prey travels faster and has to avoid colliding with the predators, while the 3 predators travel slower and have to work together to capture the prey. The rewards received by the prey and a predator sum to 0 (i.e., the prey gets a negative reward for collision, while the predators get rewarded positively), and all predators receive the same reward. The prey is also negatively rewarded if it strays away from the predefined area (a $1\times1$ unit square). This environment is general-sum because it contains 3 predators and a single prey. 

\subsection{Preprocessing}
For the MPE environments, no preprocessing was done, and default environment-parameters were used for all MPE experiments.

For the Atari environments, following the recommendations of Marlos et al. \cite{revisitingALE}, we performed the following preprocessing - reward clipping, sticky actions, frame skipping, and no-op resets. The number of steps per episode was also set to a limit of 200 for both Atari environments, as that yielded the best results in general.
Furthermore, the action spaces for both Atari environments were shrunk to their effective action spaces in order to improve learning efficiency. For Pong specifically, we also concatenated a one-hot vector of the agent's index to the observations so that independent algorithms can differentiate one from the other when parameter sharing is utilized. Further details of the preprocessing performed can be found in Appendix \ref{appendix:preprocessing}.

\subsection{Implementation}
Implementations of all algorithms were based on open-sourced libraries/reference implementations. Default hyperparameters were used for all algorithms, and no hyperparameter tuning was performed. Details of implementations, alongside their hyperparameters, can be found in Appendix \ref{appendix:implementation}.  

All experiments were performed across 5 different seeds. Parameter sharing was utilized for all algorithms throughout the experiments for all environments with homogeneous state and action spaces. For multi-agent algorithms that perform centralized training (e.g., QMIX, COMA, MADDPG etc.), the global states were represented by the concatenation of all agents' local observations. We also used the 128-byte Atari RAM as state inputs, rather than visual observations. This allows the algorithms to focus their learning on control rather than on both control and perception, improving learning efficiency. 

\section{Experiment Results}

In this section, we highlight the experiments performed on the four multi-agent environments (i.e., Simple Reference, Space Invaders, Pong and Simple Tag), and provide discussions about the obtained results. 

\subsection{Cooperative}

\begin{figure}
    \centering
    \begin{subfigure}{0.5\textwidth}
         \includegraphics[width=\textwidth]{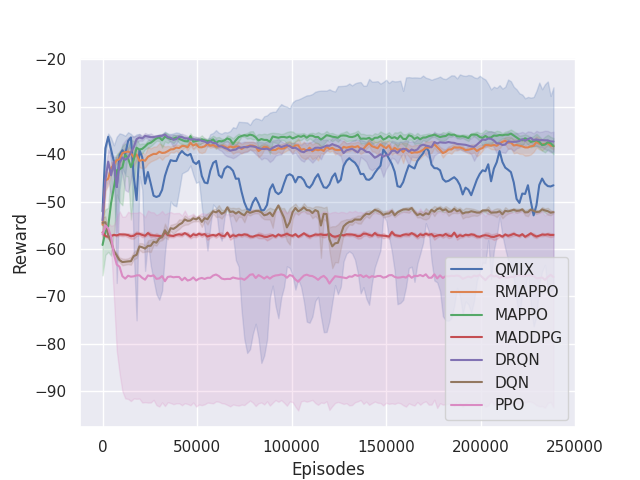}
         \caption{Simple Reference}
         \label{fig:simple_reference_results}
    \end{subfigure}
    \begin{subfigure}{0.49\textwidth}
         \includegraphics[width=\textwidth]{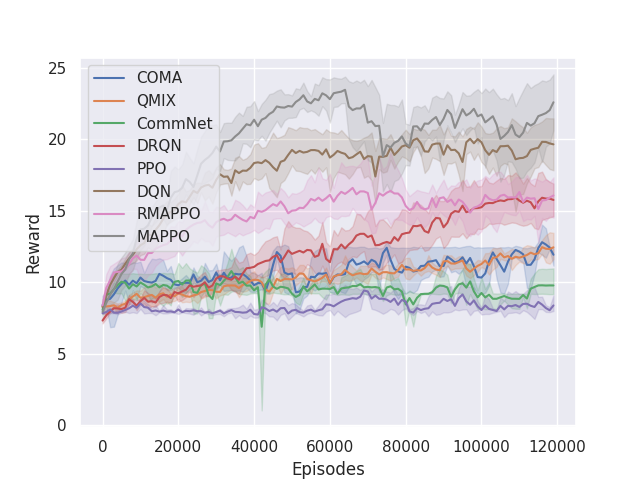}
         \caption{Space Invaders}
         \label{fig:space_invaders_results}
    \end{subfigure}
    \caption{Training curves of various algorithms in two cooperative environments. For every algorithm, the solid line represents the mean reward per episode, while the shaded region represents the 95\% confidence interval around the mean.}
    \label{fig:cooperative_envs_results}
\end{figure}

\paragraph{Simple Reference}

We ran the various algorithms on the Simple Reference environment for 240k episodes ($6 \times 10^6$ steps). From Fig.~\ref{fig:simple_reference_results}, it could be observed that all independent algorithms converged to a lower score, except for DRQN, whose recurrence allowed it to vastly outperform DQN and converge to a score on par with multi-agent algorithms. However, this trend was not observed when comparing MAPPO to its recurrent variant (i.e., RMAPPO), as MAPPO performs equally well as RMAPPO. We hypothesize that since MAPPO's centralized critic learns based on the joint observation and action of both agents, 
this minimizes the amount of partial observability of every agent, 
and allows each agent to learn to communicate with other agents effectively without recurrence.
In contrast, for independent algorithms, such as DQN, where the interactions between the agents are not explicitly learned (since all other agents are treated as part of the environment), adding recurrence could help mitigate some resulting partial observability, hence improving their performance, as described above.

\paragraph{Space Invaders}
Unlike the Simple Reference environment, the Space Invaders environment seemed to favour non-recurrent variants of algorithms (Fig.~\ref{fig:space_invaders_results}). MAPPO vastly outperformed RMAPPO, and similarly DQN outperformed DRQN. This is also likely the underlying reasoning behind the comparatively poorer performance of the multi-agent algorithms, such as QMIX, COMA and CommNet, all of which were implemented with recurrent neural networks under the CTDE scheme. 

Additionally, since there is no unit collision in the Space Invaders environment (i.e., agents can move past each other without being blocked), they do not have to coordinate between themselves to achieve a high score in the environment; a good policy can be learned solely by having agents maximize their individual rewards. This explains the strong performance that was achieved by DQN.
Also, since this is a cooperative task with both agents having identical goals, learning separate representations for individual agents is not very important; the learning of both agents assist each other. This is shown in Fig.~\ref{fig:dqn_agent_indicator_spi} in Appendix~\ref{appendix:importance-of-agent-indicator}, where the addition of an agent indicator did not yield any performance improvement for DQN on Space Invaders.

Given such circumstances, it is interesting to observe the stronger performance of MAPPO compared to the independent algorithms. By conditioning on the joint action, MAPPO's critic has full observability into the joint action that resulted in the team reward. Therefore, the observed reward is unbiased, which allows the learning process to be more efficient. In contrast, independent algorithms have to learn from a noisy team reward signal, where an agent could receive a large positive team reward even when it did nothing. This relates to the problem of credit assignment in MARL, noted in prior works \cite{hernandez2019survey}.   

\subsubsection{Multi-Agent Credit Assignment Problem in Fully Observable Settings} \label{section:effect_of_team_reward}
 
In this section, we attempt to study the effect of using a team reward signal, rather than individual reward signals on various independent and multi-agent algorithms in a fully observable environment. When team rewards are the only rewards given, these reward signals are noisy for independent algorithms because the agent, which treats every other agent as part of the environment, does not know the actions taken by other agents. This makes it difficult for independent algorithms' agents to learn how their individual actions contribute to the team reward signal. We performed the experiments on Space Invaders, in which the default agents receive individual rewards from the environment. To study the effect of the multi-agent credit assignment problem, we performed two runs per algorithm, one with team rewards only, and the other with individual rewards only (i.e., agents are rewarded independently by the environment). 

\begin{figure}
     \centering
    %  \begin{subfigure}{.3\textwidth}
    \begin{subfigure}{.48\textwidth}
         \includegraphics[width=\textwidth]{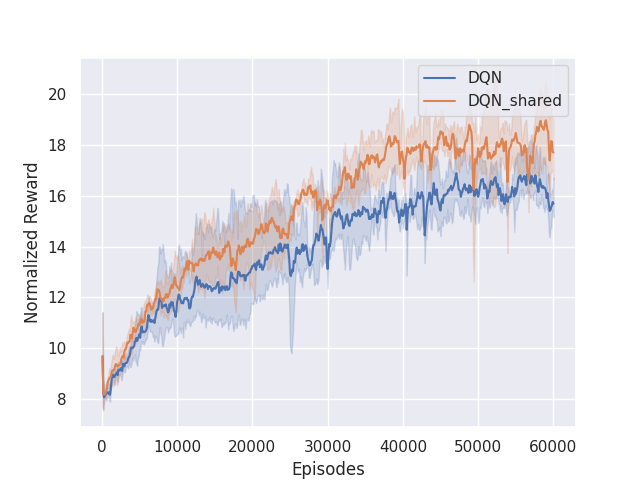}
         \caption{DQN}
         \label{fig:dqn_shared_rewards_comparison}
     \end{subfigure}
    
    %  \begin{subfigure}{.3\textwidth}
    \begin{subfigure}{.48\textwidth}
         \includegraphics[width=\textwidth]{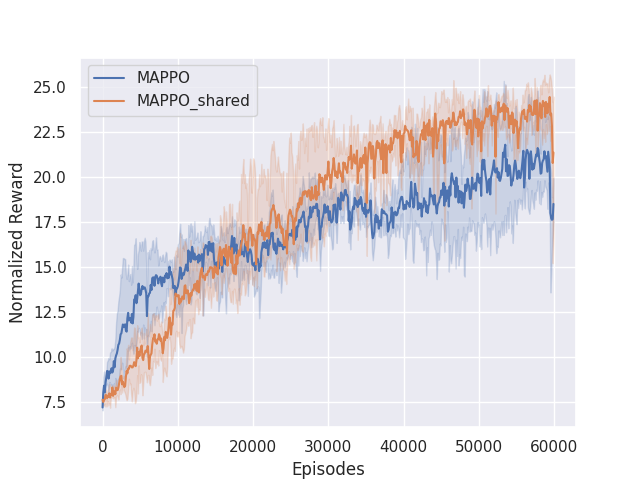}
         \caption{MAPPO}
         \label{fig:mappo_shared_rewards_comparison}
     \end{subfigure}
    %  \begin{subfigure}{.3\textwidth}
    \begin{subfigure}{.48\textwidth}
         \includegraphics[width=\textwidth]{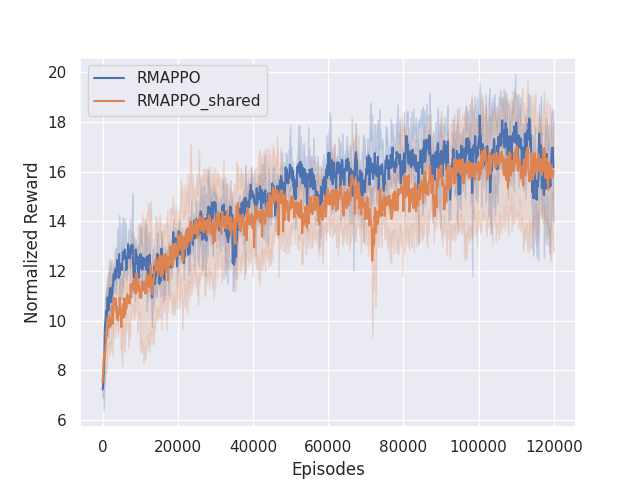}
         \caption{RMAPPO}
         \label{fig:rmappo_shared_rewards_comparison}
     \end{subfigure}
     \caption{Training curves of various algorithms in Space Invaders, comparing when individual rewards are given (blue) to when team rewards are given (orange).}
     \label{fig:comparing_shared_rewards}
 \end{figure}

For multi-agent algorithms, such as MAPPO (Fig.~\ref{fig:mappo_shared_rewards_comparison}) and RMAPPO (Fig.~\ref{fig:rmappo_shared_rewards_comparison}), 
having a team reward does not have a large effect on the performance of the algorithms. This is expected because these algorithms have critics that learn from the joint action, which allow them to implicitly learn the estimated contribution of every agent without factorization.  
 
On similar lines, regarding  independent algorithms, we observe that having team rewards instead of individual ones do not impact their performance adversely (Fig.~\ref{fig:dqn_shared_rewards_comparison}). A plausible explanation could be that since all agents receive the same reward for a given joint action, this allows the independent algorithms to correlate actions from different agents that produced similar (high) rewards.

\subsection{Competitive}

The 2-player Pong environment was used for the competitive setting. All algorithms were first trained using parameter sharing with the addition of agent indicators (the effect of which is detailed in Appendix \ref{appendix:importance-of-agent-indicator}) for 60k episodes ($1.2\times10^6$ steps), their network parameters were then saved. Since Pong is a zero-sum game, we evaluated them by putting them head-to-head against each other for 3 episodes for all possible combinations. After that, their positions were swapped, and the entire process was repeated. Swapping their positions is crucial for evaluation, because the first player (playing the right paddle) is always the serving player, therefore the first player always has an advantage over the second player (which plays the left paddle). This advantage is further exacerbated because the winning side always gets to serve subsequent openings. The entire evaluation process was repeated across all 5 seeds.

\begin{figure}
    \centering
    % \begin{subfigure}[t]{.3\textwidth}
    \begin{subfigure}[t]{.45\textwidth}
        \includegraphics[width=\textwidth]{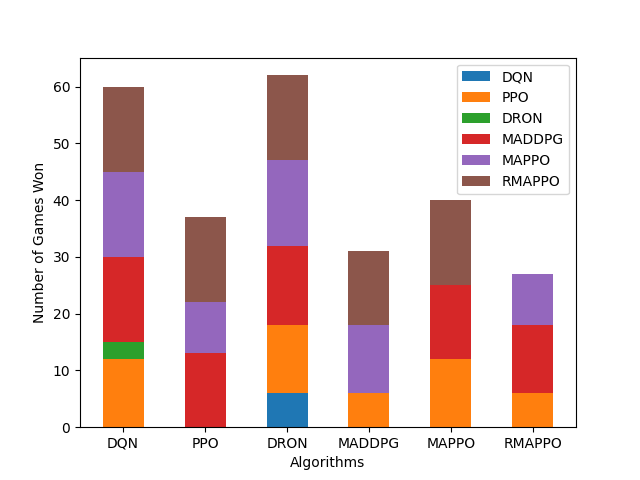}
        \caption{Number of games won as the first player}
        \label{fig:num_games_won_as_first}
    \end{subfigure}
    % \begin{subfigure}[t]{.3\textwidth}
    \begin{subfigure}[t]{.45\textwidth}
        \includegraphics[width=\textwidth]{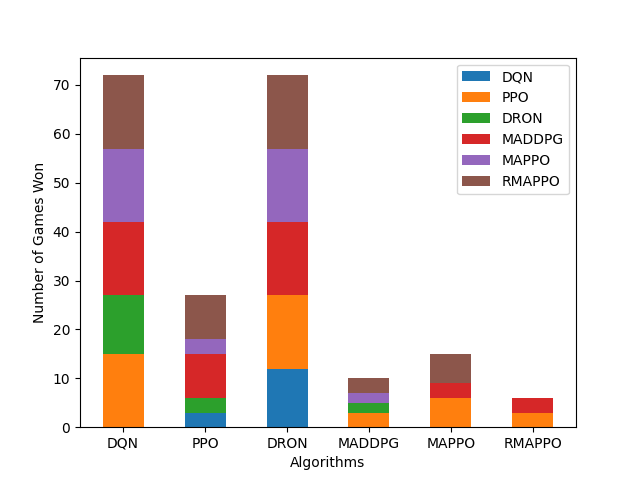}
        \caption{Number of games won as the second player}
        \label{fig:num_games_won_as_second}
    \end{subfigure}
    \begin{subfigure}[t]{.45\textwidth}
        \includegraphics[width=\textwidth]{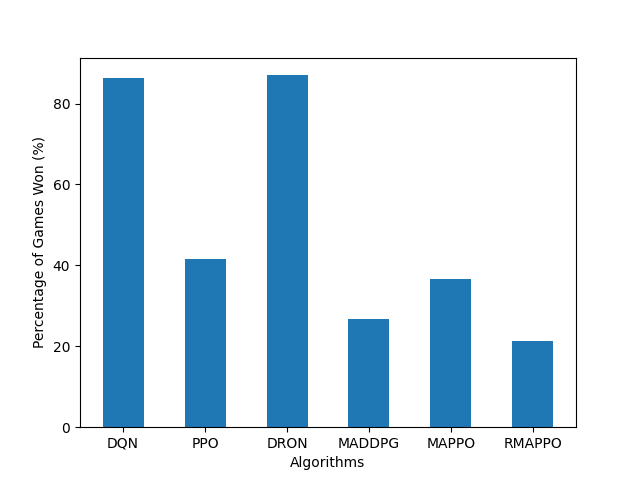}
        \caption{Overall win rate percentage}
        \label{fig:winrate_in_pong}
    \end{subfigure}
    
    \caption{Performance of various algorithms when playing against other algorithms in Pong.}
    \label{fig:pong_results}
\end{figure}

From the stacked bar charts shown in Fig.~\ref{fig:pong_results}, a similar trend across the number of games won as the first and second player can be observed (Fig.~\ref{fig:num_games_won_as_first} and \ref{fig:num_games_won_as_second}). DRON is consistently the best player, closely followed by DQN. Both of these algorithms were also the only algorithms to have a win rate of greater than 50\% for the games they have played (Fig.~\ref{fig:winrate_in_pong}). 

An interesting observation that can be made is the strong performance of independent algorithms, compared to other multi-agent algorithms. Since Pong is fully observable, critics that learn based on the joint observation of both agents do not necessarily provide any new information. Furthermore, since Pong is a highly reactive environment, an agent can learn a good policy solely by understanding how to position its paddle according to the trajectory of the ball (towards the agent). While learning on the joint action could allow agents to learn to better predict the incoming trajectory of the ball, it can be observed that the additional layer of complexity causes the sample efficiency to decrease and only yields diminishing returns. 
In addition to the above factors, it is possible that parameter sharing benefited agents of independent algorithms by allowing them to learn better representations of both players, since they were trained to play as both players simultaneously.
Had these algorithms trained without parameter sharing, there would likely be a larger performance difference between independent algorithms and opponent modelling algorithms such as DRON. Instead of treating other agents as part of the environment, opponent modelling allows agents to adapt more quickly to the opponent's changing strategies \cite{dron}. However, the minimal improvement DRON has over DQN suggests that in the Pong environment, an agent's policy may not be significantly affected by changes in the opponent agent's policy (i.e., individual agents can play the same way regardless of how their opponent played).

\subsection{Mixed}
\begin{figure}
    \centering
    \begin{subfigure}[t]{.48\textwidth}
        \includegraphics[width=\textwidth]{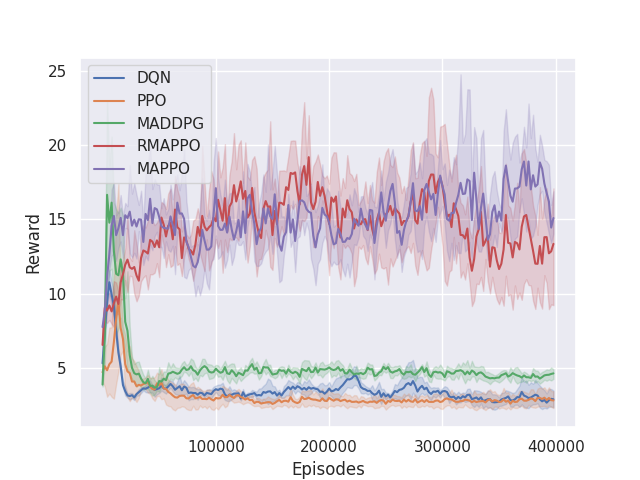}
        \caption{Reward of a predator; all predators obtain the same reward}
        \label{fig:predator_simple_tag}
    \end{subfigure}
    \begin{subfigure}[t]{.48\textwidth}
        \includegraphics[width=\textwidth]{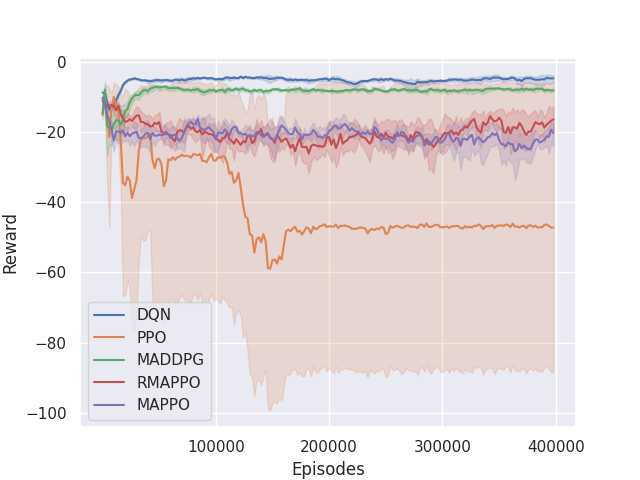}
        \caption{Reward of the prey}
        \label{fig:prey_simple_tag}
    \end{subfigure}
    
    \caption{Training curves of various algorithms in the Simple Tag, a Predator-Prey environment}
    \label{fig:simple_tag_results}
\end{figure}

In the Simple Tag (i.e., Predator-Prey) environment, the predators are incentivized to cooperate together to trap the prey, while the prey is incentivized to dodge the predators while staying within a predefined area. For our method of evaluation, we plot the training curves of the prey (Fig.~\ref{fig:prey_simple_tag}), and one of the predators (Fig.~\ref{fig:predator_simple_tag}), since all predators receive the same reward. 
Since the observation and action spaces differ between the predators and the prey, none of the agents have their parameters shared. We chose not to share the parameters of the predators to ensure that bias towards the predators was not introduced (since they would have 3 times the amount of data to work with compared to the prey). 

In the case of DQN, the prey successfully learned to minimize the number of collisions with the predators, which can be observed by the strong performance achieved by the prey (Fig.~\ref{fig:prey_simple_tag}). However, similar to PPO, since the predators were trained completely independently (i.e., their parameters were not shared), they did not manage to learn how to cooperate with one another to capture the prey (Fig.~\ref{fig:predator_simple_tag}). It is interesting to observe that MADDPG converged to a policy similar to DQN, with the difference being that its predators have learned to cooperate better, thus getting slightly higher rewards compared to DQN's predators (Fig.~\ref{fig:predator_simple_tag}). Subsequently, as a result of the higher rewards obtained by the predators, MADDPG achieves a slightly lower score for its prey (Fig.~\ref{fig:prey_simple_tag}).

MAPPO and RMAPPO, on the other hand, learned a different strategy. As we can observe from the comparatively noisier curves obtained from their predators and preys (Fig. \ref{fig:predator_simple_tag} and \ref{fig:prey_simple_tag}), there is a constant tug-of-war between the prey and the predators - as the predators learn how to cooperate better, their scores increase, which subsequently causes the prey to learn how to dodge, decreasing the predators' scores, and vice versa. Since the predators of MAPPO and RMAPPO achieves a much higher score compared to all other algorithms, this is indicative that the predators have successfully learned to cooperate to trap the prey.

\section{Conclusion}
\paragraph{Cooperative} In the cooperative setting, for environments where individual agents have full observability such as Space Invaders, we showed that independent algorithms can perform even better than certain multi-agent algorithms. Furthermore, we showed that independent algorithms are able to cope well with the multi-agent credit assignment problem in environments that are fully observable with a relatively small number of agents, and where every agent has the same task. On the other hand, in the Simple Reference environment where the need for agents to communicate induces partial observability, adding recurrence allowed independent algorithms to perform as well as other multi-agent algorithms. 
We also discussed the significance of learning on the joint observation and action, rather than individual ones, and showed that MAPPO performs as well as DRQN in the Simple Reference environment, without the need for an RNN. Moreover, in Space Invaders, MAPPO was able to consistently achieve the highest score amongst all other algorithms.

\paragraph{Competitive} In the Pong environment, we saw that DRON and DQN were able to outperform all other algorithms. We argued that this is due to the fully observable nature of the Pong environment, in addition to the diminishing returns that learning from joint actions could yield. Furthermore, we showed that with the use of agent indicators, independent algorithms were able to learn robust policies for both competing agents using parameter sharing. 

\paragraph{Mixed} In the Predator-Prey environment, we saw that since there were no parameter sharing to induce cooperation, predators from independent algorithms were unable to learn how to cooperate with each other to capture the prey. Conversely, in DQN we saw that its prey was able to achieve the highest score consistently, showing that the prey has learned to dodge the predators effectively while staying within the predefined area. Interestingly, we also saw how MADDPG's training curve for its predators and prey shows resemblance to that of DQN, suggesting that it also faced difficulties in learning strategies for the predators to coordinate and capture the prey. MAPPO and RMAPPO, on the other hand, were the only algorithms that managed to achieve high scores for their predators, suggesting that their predators have learned how to collaborate with each other to hunt the prey. The noisiness of their graphs suggest that there is a constant tug-of-war between the prey and the predators, as one tries to outsmart the other. 

\section{Future Work}
In this section, we highlight some future work that could potentially bring more insights into having a broader understanding of dealing with non-stationarity and partial observability for independent algorithms, both of which are common in the multi-agent setting. In the Space Invaders environment, we observed that independent algorithms were able to learn well with just a team reward. Future work could be done to determine if this was only the case for fully observable environments, or under what conditions would independent algorithms still be able to cope with the multi-agent credit assignment problem. It would also be interesting to study the performance of non-recurrent variants of multi-agent algorithms such as QMIX and COMA in fully observable environments. Since the experiments performed in this paper only included fully-observable competitive and mixed environments, future work can also include a more diverse set of environments, such as partially observable competitive and mixed environments.

\AtNextBibliography{\small}
\printbibliography[heading=bibintoc, title={References}]

\newpage
\appendix
\section{Preprocessing} \label{appendix:preprocessing}
In this section, we detail the preprocessing techniques used on the multi-agent Atari (RAM-state) environments. In our experiments, the four main preprocessing techniques performed are reward clipping, sticky actions, frame-skipping and no-op resets. Reward clipping ensures that the rewards at every timestep are clipped between the range of [-1, 1]. Sticky actions with a probability of 0.25 are used to inject stochasticity into the environment, and to enhance the robustness of the learned agents. We observed empirically that the addition of sticky actions and reward clipping did not have a significant effect on the learning process. On the other hand, adding a frame-skipping of 4 gave a significant improvement to the empirical performance of the algorithms. This is likely due to the increase in simulation-speed (e.g., the five steps that the agent takes is equivalent to experiencing 20 actual frames). Another subtle side-effect of this preprocessing is that unlike in environments such as MPE where feedback (rewards) are given immediately, reward signals are delayed in games with projectiles, such as Space Invaders -- where the laser beam shot by an agent might only hit an enemy after a number of steps. By performing frame-skipping, the effect of delayed rewards is greatly reduced (rewards received by the agent during the $x$ skipped frames are summed up, so they are not lost), simplifying the temporal credit-assignment problem.

Since our implementation truncates every episode to be 200-steps long, we perform a series of no-op actions at the start of every episode (these actions do not count into the step-limit per episode). Unlike in the DQN paper where initial no-ops were used to introduce randomness into the environment \cite{mnih2015dqn}, the purpose of using no-ops in this case was to allow us to skip ahead a set number of frames at the start of every game. Space Invaders, for example, has a stall state for the first $\sim130$ frames of every game, likely meant for human players to prepare for the start of the game. Removing these frames helped increase the learning efficiency for our agents. For Space Invaders and Pong, we perform no-op for the first 130 and 60 frames, respectively. 

All preprocessing were performed using the SuperSuit library \cite{SuperSuit}.

\section{Importance of Agent Indicator} \label{appendix:importance-of-agent-indicator}
In this section we list some interesting findings regarding the addition of agent indicators to independent algorithms when performing parameter sharing. 

\begin{figure}
    \centering
    \begin{subfigure}[t]{.3\textwidth}
        \includegraphics[width=\textwidth]{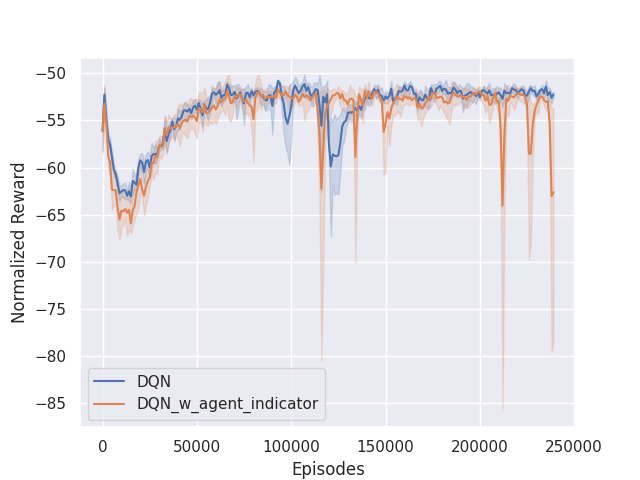}
        \caption{Simple Reference}
        \label{fig:dqn_agent_indicator_reference}
    \end{subfigure}
    \qquad
    \begin{subfigure}[t]{.3\textwidth}
        \includegraphics[width=\textwidth]{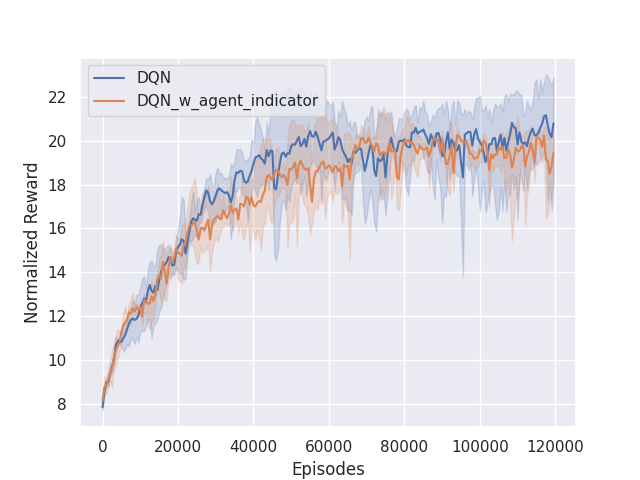}
        \caption{Space Invaders}
        \label{fig:dqn_agent_indicator_spi}
    \end{subfigure}
    \caption{Comparing DQN with (blue) and without (orange) agent indicators in Simple Reference and Space Invaders environment}
    \label{fig:dqn_agent_indicator_comparison}
\end{figure}

Interestingly, in both cooperative environments, there is no noticeable improvement in the performance of independent algorithms when an agent indicator was added (Fig.~\ref{fig:dqn_agent_indicator_comparison}). As was previously discussed in the experimental results section, in the case of Space Invaders, this is likely due to the similarities of both agents in terms of their tasks, and their representations (i.e., both agents have the same tasks and maximize the same objectives), therefore there is less of a need to distinguish between either agent. On the other hand, for the Simple Reference environment, it is very likely that the agent indicators did not make a noticeable difference because of the partially observable nature of the environment; adding recurrence would result it a much more significant difference instead.

\begin{figure}
    \centering
    \begin{subfigure}[t]{.3\textwidth}
        \includegraphics[width=\textwidth]{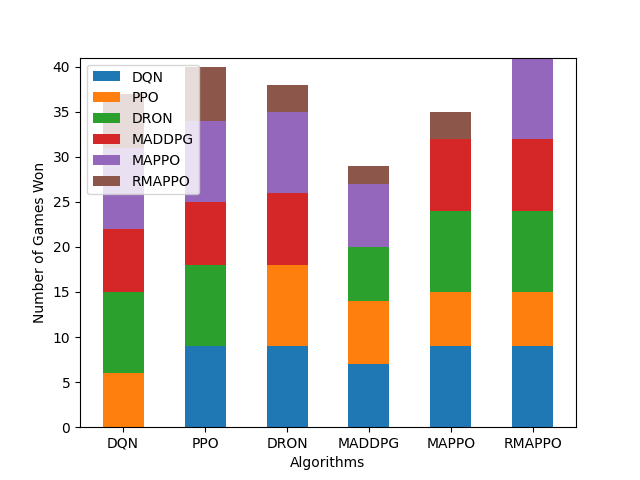}
        \caption{Number of games won as the first player}
        \label{fig:num_games_won_as_first_without_agent_indicator}
    \end{subfigure}
    \begin{subfigure}[t]{.3\textwidth}
        \includegraphics[width=\textwidth]{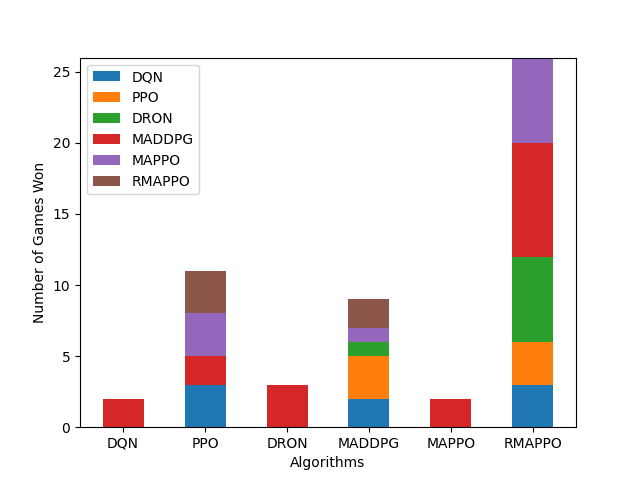}
        \caption{Number of games won as the second player}
        \label{fig:num_games_won_as_second_without_agent_indicator}
    \end{subfigure}
    \begin{subfigure}[t]{.3\textwidth}
        \includegraphics[width=\textwidth]{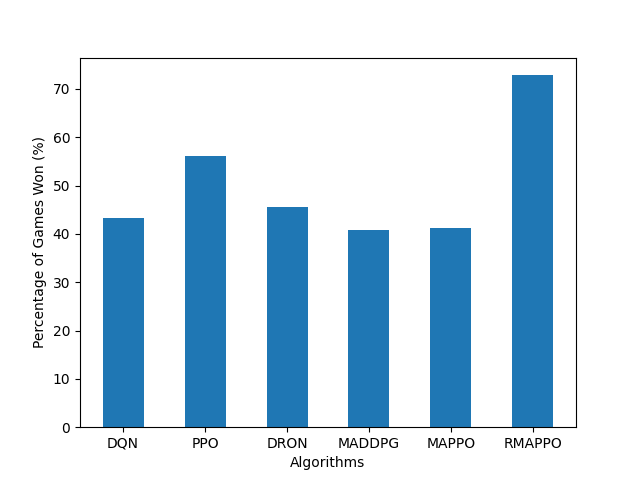}
        \caption{Overall win rate percentage}
        \label{fig:winrate_in_pong_without_agent_indicator}
    \end{subfigure}
    \caption{Putting algorithms against each other in Pong without agent indicators across 3 seeds}
    \label{fig:pong_results_without_agent_indicator}
\end{figure}

Conversely, for the Pong environment, even though it is also fully observable (akin to Space Invaders), the representation of both agents are not interchangeable. Utilizing parameter sharing without agent indicators, all algorithms struggled to learn due to the inability to tell which paddle was it controlling at every timestep. The only exception was RMAPPO (Fig.~\ref{fig:pong_results_without_agent_indicator}), which was able to condition on the sequence of previous observations to figure out which paddle was it controlling.

\section{Implementation Details} \label{appendix:implementation}
The following list contains the sources of the reference implementations for the various algorithms:
\begin{itemize}
    \item Implementation of DQN and DRON were based on the Machin library \cite{machin}. 
    \item Implementation of independent PPO was based on Stable Baselines3 \cite{stable-baselines3}. 
    \item Implementation of DRQN, QMIX, COMA and CommNet came from a popular public repository by the name of StarCraft \cite{starry-sky-starcraft}. 
    \item Implementation of MADDPG came from the original code implementation \cite{lowe_maddpg}.
    \item Implementation of MAPPO came from the original code implementation \cite{yu2021mappo}. 
\end{itemize}

For both DQN and DRON, the underlying DQN implementations include Double DQN \cite{vanhasselt2015deep}, the dueling architecture \cite{wang2016dueling} and priority experience replay buffer \cite{schaul2016prioritized}. On the other hand, the implementation of DRQN did not use any of the aforementioned add-ons. For PPO and MAPPO, 4 parallel workers were used for all environments with homogeneous state and action spaces. 

\subsection{Hyperparameters}
In this section, we specify the hyperparameters used for all algorithms used throughout the experiments. 
\begin{table}[H]
    \caption{Hyperparameters for DQN and DRON}
    \label{table:hyperparams_dqn_dron}
    \centering
    \begin{tabular}{ll}
        \toprule
        Hyperparameter & Value\\ 
        \midrule
        fully-connected layer dimensions & 512$\times$256\\ 
        % \hline
        optimizer & Adam \\
        % \hline
        learning rate & 0.001 \\
        % \hline
        discount factor & 0.99\\
        % \hline
        replay buffer size & $1\times 10^6$\\
        % \hline
        batch size & 256 \\
        % \hline
        loss function & MSE \\
        % \hline
        initial epsilon &  1 \\
        % \hline
        epsilon decay rate & 0.9999\\
        % \hline
        double & True \\
        % \hline
        dueling & True \\
        % \hline
        priority & True\\
        \bottomrule
    \end{tabular}
\end{table}

\begin{table}[H]
    \caption{Hyperparameters for PPO}
    \label{table:hyperparams_ppo}
    \centering
    \begin{tabular}{ll}
        \toprule
        Hyperparameter & Value\\ 
        \midrule
        fully-connected layer dimensions & 64$\times$64\\ 
        % \hline
        number of environments & 4\\
        % \hline
        optimizer & Adam \\
        % \hline
        number of steps & episode length\\
        % \hline
        number of epochs & 10\\
        % \hline
        minibatch size & episode length*\# of agents*4 \\
        % \hline
        discount factor & 0.99\\
        % \hline
        GAE lambda & 0.95\\
        % \hline
        learning rate & 0.0007\\
        % \hline
        value function coefficient & 0.5 \\
        % \hline
        clip & 0.2\\
        entropy & 0.01 \\
        \bottomrule
    \end{tabular}
\end{table}

\begin{table}[H]
    \caption{Hyperparameters for MADDPG}
    \label{table:hyperparams_maddpg}
    \centering
    \begin{tabular}{ll}
        \toprule
        Hyperparameter & Value\\ 
        \midrule
        fully-connected layer dimensions & 64$\times$64\\ 
        % \hline
        optimizer & Adam \\
        % \hline
        learning rate & 0.01 \\
        % \hline
        discount factor & 0.95\\
        % \hline
        replay buffer size & $1\times 10^6$\\
        % \hline
        batch size & 1024 \\
        % \hline
        critic loss function & MSE \\
        % \hline
        gradient clip norm & 0.5\\
        \bottomrule
    \end{tabular}
\end{table}

\begin{table}[H]
    \caption{Hyperparameters for MAPPO and RMAPPO}
    \label{table:hyperparams_mappo_rmappo}
    \centering
    \begin{tabular}{ll}
        \toprule
        Hyperparameter & Value\\ 
        \midrule
        fully-connected layer dimensions & 64$\times$64\\ 
        % \hline
        number of environments & 4\\
        % \hline
        optimizer & Adam \\
        % \hline
        number of epochs & 10\\
        % \hline
        minibatch size & 1600\\
        % \hline
        discount factor & 0.99\\
        % \hline
        GAE lambda & 0.95\\
        % \hline
        learning rate & 0.0007\\
        % \hline
        value function coefficient & 0.5 \\
        % \hline
        clip & 0.2\\
        % \hline
        entropy & 0.01 \\
        \midrule
        RMAPPO-specific Hyperparameters & \\
        \midrule
        number of GRU layers & 1\\
        % \hline
        hidden state dimension & 64\\
        \bottomrule
    \end{tabular}
\end{table}

\begin{table}[H]
    \caption{Hyperparameters for COMA, QMIX, DRQN and CommNet}
    \label{table:hyperparams_coma_qmix_drqn_commnet}
    \centering
    \begin{tabular}{ll}
        \toprule
        Hyperparameter & Value\\ 
        \midrule
        discount factor & 0.99\\
        optimizer & RMSProp\\
        number of GRU layers & 1\\
        hidden state dimension & 64\\
        gradient clip norm & 10\\
        batch size & 256\\
        \midrule
        COMA-specific Hyperparameters & \\
        \midrule
        critic (fully-connected) dimension & 128\\
        actor learning rate & 0.0004\\
        critic learning rate & 0.003\\
        
        \midrule
        QMIX/DRQN-specific Hyperparameters & \\
        \midrule
        hypernetwork dimension & 64\\
        learning rate & 0.0005\\
        epsilon & linear decay from 1 to 0.05\\
        buffer size & 1$\times 10^6$\\
        \bottomrule
    \end{tabular}
\end{table}

\end{document}